\input{aipcheck}

\documentclass[
    ,final            % use final for the camera ready runs
  ]
  {aipproc}

\layoutstyle{6x9}
\usepackage{epic,amsmath,graphicx}

\begin{document}

%\title{Validity of our semi-relativistic potential model for
%heavy-light systems
\title{Viable semi-relativistic quark model of heavy-light
systems
\footnote{Talk presented by T. Matsuki 
at the Workshop on
"Scalar Mesons and Related Topics (Scadron 70)" held at
Instituto Superior Tecnico (IST), Lisbon, Portugal,
February 11-16, 2008.}
}


\author{Takayuki Matsuki}
{
  address={Tokyo Kasei University,
1-18-1 Kaga, Itabashi, Tokyo 173, JAPAN
\footnote{E-mail: matsuki@tokyo-kasei.ac.jp}}
}

\author{Toshiyuki Morii}{
  address={Faculty of Human Development,
Kobe University,\\ Nada, Kobe 657-8501, JAPAN
\footnote{E-mail: morii@kobe-u.ac.jp}}
}

\author{Kazutaka Sudoh}{
  address={Institute of Particle and Nuclear Studies, 
High Energy Accelerator Research Organization, \\ 
1-1 Ooho, Tsukuba, Ibaraki 305-0801, JAPAN
\footnote{E-mail: kazutaka.sudoh@kek.jp}}
}

\begin{abstract}
The semi-relativistic quark potential model is surprisingly powerful for heavy-light systems if the bound state equation is treated correctly using $1/m_Q$ expansion with heavy quark mass $m_Q$.  We elucidate the reasons why our semi-relativistic model succeeds in predicting and reproducing all the mass
spectra of heavy-light systems so far reported, $D/D_s/B/B_s$, by reviewing and comparing recent
experimental data with the results of our model and others. Especially the mass spectra of the
so-called $D_{sJ}$, i.e., $D_{s0}^*$ and $D_{s1}'$, are successfully reproduced only by our model
but not by other models.
\end{abstract}

\maketitle

%%%%%%%%%%%%%%%%%%%%%%%%%%%%%%%%%%%%%%%%%%%%
%% MAINMATTER
%%%%%%%%%%%%%%%%%%%%%%%%%%%%%%%%%%%%%%%%%%%%

\section{Introduction}
Beginning from the discovery of the narrow meson states $D_{s0}^*(2317)$ and $D'_{s1}(2457)$
(the so-called $D_{sJ}$) by BaBar \cite{BaBar} and CLEO \cite{CLEO} in 2003, respectively, open
charm/bottom hadrons of the heavy-light systems have been discovered one after another.
Ten years before this discovery, we proposed a formulation for the semi-relativistic
potential model \cite{MM97}, based on which we have calculated mass spectra of higher states of
the heavy-light mesons.
Subsequently to the discovery of $D_{sJ}$, another set of
broad heavy mesons, $D_{0}^{*}(2308)$ and $D_{1}'(2427)$, were discovered by the
Belle collaboration \cite{Belle04}. These mesons are identified as $c\bar q$ ($q=u/d$)
excited ($\ell=1$) bound states and have the same quantum numbers, $j^P=0^+$ and
$1^+$, as $D_{sJ}$, respectively. 
The decay widths of these excited $D_{sJ}$ mesons are narrow, since the masses
are below the $DK/D^{*}K$ threshold, and hence the dominant decay modes violate the
isospin invariance, whereas the excited $D$ mesons, $D_{0}^{*}(2308)$ and $D_{1}'(2427)$,
are broad because there is no such restriction as in $D_{sJ}$ cases. More recent experiments
reported by CDF and D0 \cite{CDF_D006} found narrow $B$ and $B_s$ states with $\ell=1$,
$B_1(5720)$, $B_2^*(5745)$, $B_{s2}^*(5839)$, and $B_{s1}(5829)$. These are narrow because they
decay through the D-waves.

Utilizing our semi-relativistic potential model, we have so far solved the following problems.
\begin{enumerate}
\item Construct the formulation how to calculate the mass spectra of heavy-light systems.
\item Numerically calculate the mass spectra of these systems and compare them with the experiments.
\item Predicted mass spectra for $D_{s0}^*(2317)$, $D'_{s1}(2457)$, $D_{0}^{*}(2308)$, and
      $D_{1}'(2427)$ \cite{MM97} agrees well with the experiments.
\item Predict that $0^+$ and $1^+$ of $B_s$ are also below the threshold $BK/B^*K$. \cite{MMMS05}
\item Refurbish the calculations of \cite{MM97} and fit these with the experimental data,
  $B_1(5720)$, $B_2^*(5745)$, and $B_{s2}^*(5839)$ together with the above data. \cite{MMS07}
  This calculation predicted $M(B_{s1}')=5831$ MeV while the experiment observes it at 5829 MeV.
  \cite{CDF_D006}
\item Fit our calculations with the experimentally observed radial excitations, $n=2$ $D_s^*(2715)$ and
  $D_s^0(2860)$, and to obtain other radial excitations of $D/D_s/B/B_s$. \cite{MMS072}
\item Explain the superficially recovered global $SU(3)$ invariance among $0^+$ states of $D$ and $D_s$.
  \cite{MMS073}
\item Calculate the KM matrix elements by first calculating the Isgur-Wise functions from the wave
  functions used in computing the above mass spectra. \cite{MS07}
\end{enumerate}
Note that the difference between the experimental data of $D_{sJ}$ and the threshold $DK/D^*K$ is
only about 30 MeV but that the difference between our calculations for $B_{sJ}$ and the threshold
$BK/B^*K$
is about 200 MeV. Hence the trials to explain $D_{sJ}$ as a loosely bound $D$ and $K$ molecule can not
be applied to the case for $B_{sJ}$. Because it is hard to imagine that $D_{sJ}$ and $B_{sJ}$
have different structures, we believe that our explanation for these states as $Q\bar q$ states
is legitimate both for $D_{sJ}$ and $B_{sJ}$.

%%%%%%%%%%%%%%%%%%%%%%%%%%%%%%%%%%%%%%%%%%%%
\section{Our Semi-Relativistic potential model}
%%%%%%%%%%%%%%%%%%%%%%%%%%%%%%%%%%%%%%%%%%%%

Our formulation \cite{MM97} using the Cornell potential is to expand Hamiltonian,
energy, and wave function in terms of $1/m_Q$ and sets coupled
equations order by order. The non-trivial differential equation is obtained
in the zeroth order, which gives orthogonal set of eigenfunctions, and
quantum mechanical perturbative corrections to energy and wave functions
in higher orders are formulated. Applying the Foldy-Wouthuysen-Tani (FWT) transformation to a heavy quark and the Hamiltonian, eigenvalue equation becomes 
\begin{eqnarray}
  && H \psi_\ell = E^\ell \psi_\ell, \nonumber \\
  && \left(H_{-1} + H_0 + H_1 + \cdots \right) \left(
  \psi_{\ell 0}+\psi_{\ell 1}+\cdots \right) = % \nonumber \\
  \left(E^\ell_0+E^\ell_1+\cdots \right) \left(
  \psi_{\ell 0}+\psi_{\ell 1}+\cdots \right),
\end{eqnarray}
with the Cornell potential given by
\begin{equation*}
  S(r)=\frac{r}{a^2}+b,\quad V(r)=-\frac{4}{3}\frac{\alpha_s}{r},
\end{equation*}
where integers of subscripts and superscripts denote order in $1/m_Q$ and 
$H=H_{\rm FWT} -m_Q$ \cite{MM97}. 
The FWT transformation is not a simple non-relativistic reduction but it also includes the
effects of the negative components of the heavy quark.
We have the following expanded Hamiltonians:
\begin{eqnarray}
% H_{{\rm{FWT}}}  - m_Q  &=& H_{ - 1}  + H_0  + H_1  + H_2  \\ 
  H_{ - 1}^{ +  + }  &=&  - (1 + \beta _Q )m_Q,  \\ 
  H_{0 }^{ +  + }  &=& \vec \alpha _q  \cdot \vec p + \beta _q \left( {m_q  + S} \right) + V, \\ 
  H_{1 }^{ +  + }  &=& \frac{1}{{2m_Q }}\vec p^2  + \frac{1}{{2m_Q }}V\left[ {\left( {(\vec \alpha _q  \cdot \vec p) - i(\vec \alpha _q  \cdot \vec n)\partial _r } \right)} \right]\, % \nonumber \\
 - \frac{1}{{2m_Q }}\frac{1}{r}V\left( {\vec \alpha _q  \cdot \vec \Sigma _Q  \times \vec n} \right). % \\
%  H_2^{ +  + }  &=& \frac{1}{{2m_{_Q }^2 }}\beta _q \left( {\vec p + \frac{1}{2}\vec q} \right)^2 S +
%  \nonumber 
%  && \frac{1}{{8m_Q^2 }}\Delta V - \frac{1}{{4m_Q^2 }}\frac{1}{r}\left( {\beta _q S' + V'} \right)(\vec \Sigma _Q  \cdot \vec \ell ).
% \end{array}
\end{eqnarray}
Here superscripts $++$ mean that the matrix elements of the Hamiltonian are taken between the 
positive energy components of the heavy quarks. Negative components of the heavy quark have not
much contributions to the masses and so are the second orders in $1/m_Q$. Equation (2) gives the 
projection operator and determines the lowest order wave function which has only the positive component
of the heavy quark while the light-antiquark is treated as a fully relativistic Dirac particle,
being expressed by Eq. (3). The original Hamiltonian has the heavy quark symmetry (HQS) in the 
limit of $m_Q \rightarrow \infty$, and then this symmetry is broken by including the $1/m_Q$ 
correction terms. Actually the HQS is broken by the third term in Eq. (4), which only depends on
the quantum number $k$ that determines whether the HQS is broken or not. This term includes the
Dirac matrix $\vec\alpha_q$, which has only off-diagonal matrix elements so that there is no counter
term after the non-relativistic reduction. The chiral symmetry is broken in the first step (1),
which is included in the Hamiltonian in a certain limit as shown in Fig. 1, then the system breaks the HQS in the last step (2) in Fig. 1, which is nothing but the
{\sl \underline {hyperfine splitting}} owing to \underline {the third term of Eq. (4)}.
{\sl \underline {The dominant term}} for the mass is 
given by the recoil term, $\vec p^2/(2m_Q)$,
{\sl {\underline {the first term in Eq. (4)}}}.

\begin{center}
\begin{figure}[t]
\begin{picture}(370,90)
\setlength{\unitlength}{0.4mm}
 \thicklines
  \put(15,40){\line(1,0){60}}
  \put(130,57.5){\line(1,0){60}} \put(130,22.5){\line(1,0){60}}
  \put(245,65){\line(1,0){60}} \put(245,50){\line(1,0){60}}
  \put(245,30){\line(1,0){60}} \put(245,15){\line(1,0){60}}
 \thinlines
  \dottedline{3}(75,40)(130,57.5) \dottedline{3}(75,40)(130,22.5)
  \dottedline{3}(190,57.5)(245,65) \dottedline{3}(190,57.5)(245,50)
  \dottedline{3}(190,22.5)(245,30) \dottedline{3}(190,22.5)(245,15)
  \put(10,46){$k=\pm 1$ ($j_q=1/2$)}
  \put(0,15){$\left(\begin{array}{c}
              m_q\to0,\, S(r)\to0 \\  {\rm no}~ 1/m_Q~ {\rm corrections}
             \end{array}\right)$}
%  \put(148,62.5){$k=+1 (L=1)$} \put(148,27.5){$k=-1 (L=0)$}
  \put(130,62.5){$k=+1~(L=1)$} \put(130,27.5){$k=-1~(L=0)$}
  \put(310,62.5){$1^+$} \put(310,47.5){$0^+$}
  \put(310,27.5){$1^-$} \put(310,12.5){$0^-$}
  \put(132,7.5){($m_q\neq 0,\, S\neq0$)} \put(242,0){($1/m_Q$ corrections)}
\end{picture}
\caption{Procedure how the degeneracy is resolved in our model.}
\label{mass_level}
\end{figure}
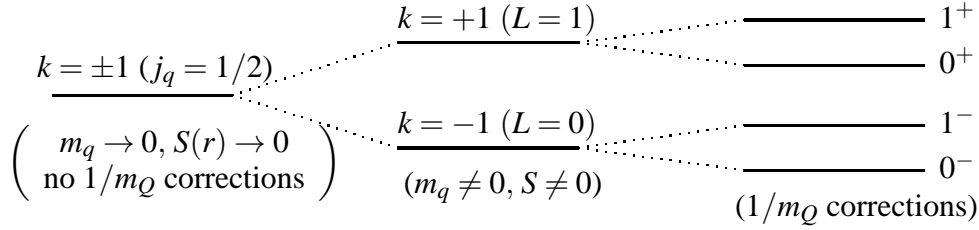
\end{center}
%

%%%%%%%%%%%%%%%%%%%%%%%%%%%%%%%%%%%%%%%%%%%%
\subsection{Mass Spectra}
%%%%%%%%%%%%%%%%%%%%%%%%%%%%%%%%%%%%%%%%%%%%

To demonstrate that how good our model calculations are, let us show in the following
Tables \ref{Dmeson}$\sim$\ref{Bsmeson} the comparison with the experimental data with the parameter set
given in Table \ref{parameter}.
\begin{table}[t!]
\caption{Optimal values of parameters.}
\label{parameter}
\centering
\begin{tabular}{lcccccccc}
\hline
\hline
%First order 
Parameters
& ~~$\alpha_s^c$ & ~~$\alpha_s^b$ & ~~$a$ (GeV$^{-1}$) & ~~$b$ (GeV) \\
& ~~0.261$\pm$0.001 & ~~0.393$\pm$0.003 & ~~1.939$\pm$0.002 & ~~0.0749$\pm$0.0020 \\
%\hline
& ~~$m_{u, d}$ (GeV) & ~~$m_s$ (GeV) & ~~$m_c$ (GeV) & ~~$m_b$ (GeV) \\
& ~~0.0112$\pm$0.0019 & ~~0.0929$\pm$0.0021 & ~~1.032$\pm$0.005 & ~~4.639$\pm$0.005 \\
\hline
\hline
\end{tabular}
\end{table}
\begin{table}[t]
\caption{$D$ meson mass spectra (units are in MeV).}
\label{Dmeson}
\begin{tabular}{@{\hspace{0.5cm}}c@{\hspace{0.5cm}}r@{\hspace{1cm}}c@{\hspace{1cm}}c@{\hspace{1cm}}l@{\hspace{1cm}}c@{\hspace{1cm}}c@{\hspace{0.5cm}}}
\hline
\hline
State ($^{2s+1}L_J$) & $k$ & $J^P$ & $M_0$ & 
\multicolumn{1}{c@{\hspace{1cm}}}{$c_1 /M_0$} & 
$M_{\rm calc}$ & $M_{\rm obs}$ \\
\hline
\multicolumn{1}{@{\hspace{1cm}}l}{$^1S_0$} 
& $-1$ & $0^-$ & 1784 & 0.476 $\times 10^{-1}$ & 1869 & 1867 \\
\multicolumn{1}{@{\hspace{1cm}}l}{$^3S_1$} 
& $-1$ & $1^-$ &  & 1.271 $\times 10^{-1}$ & 2011 & 2008 \\
\multicolumn{1}{@{\hspace{1cm}}l}{$^3P_0$} 
& $1$ & $0^+$ & 2067 & 1.046 $\times 10^{-1}$ & 2283 & 2308 \\
\multicolumn{1}{@{\hspace{1cm}}l}{$"^3P_1"$} 
& $1$ & $1^+$ &  & 1.713 $\times 10^{-1}$ & 2421 & 2427 \\
\multicolumn{1}{@{\hspace{1cm}}l}{$"^1P_1"$} 
& $-2$ & $1^+$ & 2125 & 1.415 $\times 10^{-1}$ & 2425 & 2420 \\
\multicolumn{1}{@{\hspace{1cm}}l}{$^3P_2$} 
& $-2$ & $2^+$ &  & 1.618 $\times 10^{-1}$ & 2468 & 2460 \\
\multicolumn{1}{@{\hspace{1cm}}l}{$^3D_1$} 
& $2$ & $1^-$ & 2322 & 1.894 $\times 10^{-1}$ & 2762 & $-$ \\
\multicolumn{1}{@{\hspace{1cm}}l}{$"^3D_2"$} 
& $2$ & $2^-$ &  & 2.054 $\times 10^{-1}$ & 2800 & $-$ \\
%%%%%%%%%%%%%%%%%%%%%%%%%%%
\hline
\hline
\end{tabular}
\end{table}
\begin{table}[t]
\caption{$D_s$ meson mass spectra (units are in MeV).}
\label{Dsmeson}
\begin{tabular}{@{\hspace{0.5cm}}c@{\hspace{0.5cm}}r@{\hspace{1cm}}c@{\hspace{1cm}}c@{\hspace{1cm}}l@{\hspace{1cm}}c@{\hspace{1cm}}c@{\hspace{0.5cm}}}
\hline
\hline
State ($^{2s+1}L_J$) & $k$ & $J^P$ & $M_0$ & 
\multicolumn{1}{c@{\hspace{1cm}}}{$c_1 /M_0$} & 
$M_{\rm calc}$ & $M_{\rm obs}$ \\
\hline
\multicolumn{1}{@{\hspace{1cm}}l}{$^1S_0$} 
& $-1$ & $0^-$ & 1900 & 0.352 $\times 10^{-1}$ & 1967 & 1969 \\
\multicolumn{1}{@{\hspace{1cm}}l}{$^3S_1$} 
& $-1$ & $1^-$ &  & 1.102 $\times 10^{-1}$ & 2110 & 2112 \\
\multicolumn{1}{@{\hspace{1cm}}l}{$^3P_0$} 
& $1$ & $0^+$ & 2095 & 1.101 $\times 10^{-1}$ & 2325 & 2317 \\
\multicolumn{1}{@{\hspace{1cm}}l}{$"^3P_1"$} 
& $1$ & $1^+$ &  & 1.779 $\times 10^{-1}$ & 2467 & 2460 \\
\multicolumn{1}{@{\hspace{1cm}}l}{$"^1P_1"$} 
& $-2$ & $1^+$ & 2239 & 1.274 $\times 10^{-1}$ & 2525 & 2535 \\
\multicolumn{1}{@{\hspace{1cm}}l}{$^3P_2$} 
& $-2$ & $2^+$ &  & 1.467 $\times 10^{-1}$ & 2568 & 2572 \\
\multicolumn{1}{@{\hspace{1cm}}l}{$^3D_1$} 
& $2$ & $1^-$ & 2342 & 2.032 $\times 10^{-1}$ & 2817 & $-$ \\
\multicolumn{1}{@{\hspace{1cm}}l}{$"^3D_2"$} 
& $2$ & $2^-$ &  & 2.196 $\times 10^{-1}$ & 2856 & $-$ \\
%%%%%%%%%%%%%%%%%%%%%%%%%%%
\hline
\hline
\end{tabular}
\end{table}
\begin{table}[t]
\caption{$B$ meson mass spectra (units are in MeV).}
\label{Bmeson}
\begin{tabular}{@{\hspace{0.5cm}}c@{\hspace{0.5cm}}r@{\hspace{1cm}}c@{\hspace{1cm}}c@{\hspace{1cm}}l@{\hspace{1cm}}c@{\hspace{1cm}}c@{\hspace{0.5cm}}}
\hline
\hline
State ($^{2s+1}L_J$) & $k$ & $J^P$ & $M_0$ & 
\multicolumn{1}{c@{\hspace{1cm}}}{$c_1 /M_0$} & 
$M_{\rm calc}$ & $M_{\rm obs}$ \\
\hline
\multicolumn{1}{@{\hspace{1cm}}l}{$^1S_0$} 
& $-1$ & $0^-$ & 5277 & $-0.161$ $\times 10^{-2}$ & 5270 & 5279 \\
\multicolumn{1}{@{\hspace{1cm}}l}{$^3S_1$} 
& $-1$ & $1^-$ &  & 0.981 $\times 10^{-2}$ & 5329 & 5325 \\
\multicolumn{1}{@{\hspace{1cm}}l}{$^3P_0$} 
& $1$ & $0^+$ & 5570 & 0.401 $\times 10^{-2}$ & 5592 & $-$ \\
\multicolumn{1}{@{\hspace{1cm}}l}{$"^3P_1"$} 
& $1$ & $1^+$ &  & 1.412 $\times 10^{-2}$ & 5649 & $-$ \\
\multicolumn{1}{@{\hspace{1cm}}l}{$"^1P_1"$} 
& $-2$ & $1^+$ & 5660 & 1.069 $\times 10^{-2}$ & 5720 & 5720 \\
\multicolumn{1}{@{\hspace{1cm}}l}{$^3P_2$} 
& $-2$ & $2^+$ &  & 1.364 $\times 10^{-2}$ & 5737 & 5745 \\
\multicolumn{1}{@{\hspace{1cm}}l}{$^3D_1$} 
& $2$ & $1^-$ & 5736 & 2.203 $\times 10^{-1}$ & 6999 & $-$ \\
\multicolumn{1}{@{\hspace{1cm}}l}{$"^3D_2"$} 
& $2$ & $2^-$ &  & 1.430 $\times 10^{-1}$ & 6556 & $-$ \\
%%%%%%%%%%%%%%%%%%%%%%%%%%%
\hline
\hline
\end{tabular}
\end{table}
\begin{table}[t]
\caption{$B_s$ meson mass spectra (units are in MeV).}
\label{Bsmeson}
\begin{tabular}{@{\hspace{0.5cm}}c@{\hspace{0.5cm}}r@{\hspace{1cm}}c@{\hspace{1cm}}c@{\hspace{1cm}}l@{\hspace{1cm}}c@{\hspace{1cm}}c@{\hspace{0.5cm}}}
\hline
\hline
State ($^{2s+1}L_J$) & $k$ & $J^P$ & $M_0$ & 
\multicolumn{1}{c@{\hspace{1cm}}}{$c_1 /M_0$} & 
$M_{\rm calc}$ & $M_{\rm obs}$ \\
\hline
\multicolumn{1}{@{\hspace{1cm}}l}{$^1S_0$} 
& $-1$ & $0^-$ & 5394 & $-0.302$ $\times 10^{-2}$ & 5378 & 5369 \\
\multicolumn{1}{@{\hspace{1cm}}l}{$^3S_1$} 
& $-1$ & $1^-$ &  & 0.853 $\times 10^{-2}$ & 5440 & $-$ \\
\multicolumn{1}{@{\hspace{1cm}}l}{$^3P_0$} 
& $1$ & $0^+$ & 5598 & 0.350 $\times 10^{-2}$ & 5617 & $-$ \\
\multicolumn{1}{@{\hspace{1cm}}l}{$"^3P_1"$} 
& $1$ & $1^+$ &  & 1.498 $\times 10^{-2}$ & 5682 & $-$ \\
\multicolumn{1}{@{\hspace{1cm}}l}{$"^1P_1"$} 
& $-2$ & $1^+$ & 5775 & 0.978 $\times 10^{-2}$ & 5831 & 5829 \\
\multicolumn{1}{@{\hspace{1cm}}l}{$^3P_2$} 
& $-2$ & $2^+$ &  & 1.263 $\times 10^{-2}$ & 5847 & 5839 \\
\multicolumn{1}{@{\hspace{1cm}}l}{$^3D_1$} 
& $2$ & $1^-$ & 5875 & 2.949 $\times 10^{-2}$ & 6048 & $-$ \\
\multicolumn{1}{@{\hspace{1cm}}l}{$"^3D_2"$} 
& $2$ & $2^-$ &  & 0.564 $\times 10^{-2}$ & 5908 & $-$ \\
%%%%%%%%%%%%%%%%%%%%%%%%%%%
\hline
\hline
\end{tabular}
\end{table}
In Tables \ref{Dmeson} $\sim$ \ref{Bsmeson}, $J^P$ stands for the total spin and parity, $M_0$ the lowest
degenerate mass, $c_1/M_0$ the first order correction, $M_{\rm calc}$ calculated value of mass, and
$M_{\rm obs}$ observed mass.
The calculated masses, $M_{\rm calc}$, are within one percent of accuracy compared with the observed masses,
$M_{\rm obs}$, $k$ the quantum number of the operator
$-\beta_{q} \left( \vec \Sigma_{q} \cdot \vec L + 1 \right)$, which denotes the degenerate states.
In the calculations, we have used values of parameters listed in Table 1.

%%%%%%%%%%%%%%%%%%%%%%%%%%%%%%%%%%%%%%%%%%%%
\subsection{Comparison with Other Models}
%%%%%%%%%%%%%%%%%%%%%%%%%%%%%%%%%%%%%%%%%%%%

After observing that our model nicely succeeds in predicting and/or reproducing the experimental 
data for the heavy-light mesons, we should clarify the reason why our model well works while 
others do not. Especially the other models have trouble to generate masses for the $0^+$ and
$1^+$ states of $D_s$. We will give Table \ref{comparison} which qualitatively describes the 
differences between our model and others.
\begin{table}[t]
\caption{Comparison ours with other models.
QPM means quark potential model}
\label{comparison}
\begin{tabular}{@{\hspace{0.5cm}}c@{\hspace{0.5cm}}r@{\hspace{0.5cm}}c@{\hspace{0.5cm}}c@{\hspace{0.5cm}}l@{\hspace{0.5cm}}}
\hline
\hline
Method & Authors & $D_{sJ}$ & Successful? & $m_q$  \\
\hline
\multicolumn{1}{@{\hspace{0.5cm}}l}{Semirelativistic QPM}
& T. M. et al.\cite{MM97,MMS07} & 2.339, 2.487 GeV & OK & current  \\
\multicolumn{1}{@{\hspace{0.5cm}}l}{Conventional QPM}
& Godfrey et al.\cite{GIK} & 2.48, 2.55 & No & constituent  \\
\multicolumn{1}{@{\hspace{0.5cm}}l}{BS eq. $\sim$ ours}
& Zeng et al.\cite{ZOR} & 2.38, 2.51 & No & constituent  \\
\multicolumn{1}{@{\hspace{0.5cm}}l}{Another QPM} 
& Ebert et al.\cite{EGF} & 2.463, 2.535 & No & constituent  \\
\multicolumn{1}{@{\hspace{0.5cm}}l}{$DK$ Molecule}
& Barnes et al.\cite{BCL} & $-$ & ? & N/A  \\
\multicolumn{1}{@{\hspace{0.5cm}}l}{Coupled Channel}
& Beveren et al.\cite{BRHK} & 2.28 (2.320) & OK & N/A  \\
\multicolumn{1}{@{\hspace{0.5cm}}l}{tetraquark}
& Cheng et al.\cite{CHT} & $-$ & ? & N/A  \\
%%%%%%%%%%%%%%%%%%%%%%%%%%%
\hline
\multicolumn{1}{@{\hspace{0.5cm}}l}{Observed} 
& \cite{BaBar, CLEO} & 2.317, 2.460 & $-$ & \ $-$ \\
\multicolumn{1}{@{\hspace{0.5cm}}l}{$D+K/D^*+K$} 
& $-$ & 2.367, 2.505 & $-$ & $-$  \\
%%%%%%%%%%%%%%%%%%%%%%%%%%%
\hline
\hline
\end{tabular}
\end{table}
As one can see in Table \ref{comparison}, only successful quark potential model to reproduce 
masses of $D_{sJ}$ is our semirelativistic model. A coupled channel method is also successful
but the physical meaning remains obscure in that the authors of \cite{BRHK} do not take into
account all the channels.

The BS equation is proposed by Zeng, Van Orden, and Roberts \cite{ZOR} to describe the heavy-light 
system, which is
similar to ours except that they neglect the negative components of the heavy quark.
Their numerical calculations give values higher than $DK/D^*K$ thresholds
and use constituent quark masses.
The differences between ours and theirs are i) whether the light quark masses $m_q$ are small
or not, i.e., current or constituent quark masses, ii) whether the negative components of
the heavy quark are taken into account or not. We adopt the current quark masses, $m_u=m_d=11.2$
and $m_s=92.9$ MeV while they adopt $m_u=m_d=248$ and $m_s=400$ MeV. We take into account the
negative components of the heavy quark which contribute to the second order calculations in
$1/m_Q$, while the paper \cite{ZOR} takes into account the second orders coming from only the
positive components of the heavy quark. Considering our successful calculations, we believe that if they \cite{ZOR} adopt the
current quark masses, then they would obtain the correct mass values for $D_{sJ}$ by adjusting
parameters. How the light quark mass affects the spectra can be seen in Figure 1 of Ref. 
\cite{KKMKM}, in which paper the average $D$ meson mass of $D$ and $D^{*}$ is calculated
by varying the $c$ quark mass  and by taking two values of the light quark mass,
$m_u=10$ and $336$ MeV. Even though the potential form is different from ours, this figure
shows that the value of the light quark mass is important to determine the spectra of the
heavy-light system. It turns out that only the case of $m_u=10$ MeV, i.e., current quark
mass, can fit with the experiments for the heavy-light system.
% as you can see from Figure 1 of Ref. \cite{KKMKM}.
%
%\begin{figure}[t]
%\includegraphics[scale=0.6,clip]{fig1.eps}
%\caption{Plot of $D$ meson mass by varying $m_u$. (a) for $m_u=10$ MeV and (b) for $m_u=336$
%MeV. \cite{KKMKM}}
%\label{fig2}
%\end{figure}

In the last two rows of Table \ref{comparison}, we list the experiments and the sum of
masses of $D/D^*$ and $K$ so that one can see how far calculated ones are away from
the experiments and $D/D^*K$ thresholds.

%%%%%%%%%%%%%%%%%%%%%%%%%%%%%%%%%%%%%%%%%%%%%%%%
%% BACKMATTER
%%%%%%%%%%%%%%%%%%%%%%%%%%%%%%%%%%%%%%%%%%%%%%%%

%%%%%%%%%%%%%%%%%% reference %%%%%%%%%%%%%%%%%%%
\def\Journal#1#2#3#4{{#1} {\bf #2}, #3 (#4)}
\def\NIM{Nucl. Instrum. Methods}
\def\NIMA{Nucl. Instrum. Methods A}
\def\NPB{Nucl. Phys. B}
\def\PLB{Phys. Lett. B}
\def\PRL{Phys. Rev. Lett.}
\def\PRD{Phys. Rev. D}
\def\PTP{Prog. Theor. Phys.}
\def\ZPC{Z. Phys. C}
\def\EPJC{Eur. Phys. J. C}
\def\EPJA{Eur. Phys. J. A}
\def\PR{Phys. Rept.}
\def\IJM{Int. J. Mod. Phys. A}

%\begin{thebibliography}{9}

\end{document}